*Review*

# A Survey of Distributed Certificate Authorities in MANETs

**Junaid Chaudhry[1, *], Kashif Saleem[2], Paul Haskell-Dowland[3], Mahdi H. Miraz[4]**

[1]College of Security and Intelligence, Embry-Riddle Aeronautical University, USA
chaudhrj@erau.edu
[2]Center of Excellence in Information Assurance (CoEIA), King Saud University, Saudi Arabia.
ksaleem@ksu.edu.sa
[3]School of Science, Edith Cowan University (ECU), Australia
p.haskelldowland@ecu.edu.au
[4]The Chinese University of Hong Kong (CUHK), Sha Tin, Hong Kong
m.miraz@cuhk.edu.hk
*Correspondence: chaudhrj@erau.edu

**Abstract:** A Certificate Authority (CA) provides the critical authentication and security services for Public Key Infrastructure (PKI) which are used for the Internet and wired networks. In MANETs (wireless and ad hoc) there is an inability to offer a centralized CA to provide these security services. Recent research has looked to facilitate the use of CAs within MANETs through the use of a Distributed Certificate Authority (DCA) for wireless and ad hoc networks. This paper presents a number of different types of DCA protocols and categorizes them into groups based on their factors and specifications. The paper concludes by proposing the best DCA security services in terms of performance and level of security.

## 1. Introduction

MANETs consist of mobile devices that are connected together via wireless links. MANETs suffer from a number of limitations including: the lack of any centralized topology; limited speed; and, portability [1]. Therefore, because of these limitations it is complicated to implement stable and powerful networks which are not vulnerable to different types of attacks [1, 2]. Using CAs as a powerful component of Public Key Infrastructure (PKI) in MANET networks should be a good solution in securing these kinds of networks by using a trusted third party for authenticating users [3]. Unfortunately, CAs are themselves vulnerable to compromise with potential for intruders to attack and subsequently sign certificates using the node's own private key.

Although a node can be defined as a CA, there are some problems with this approach which are related to the node's existence. If the CA node is removed from the MANET it will cause the entire network to be impacted. Furthermore, it is vulnerable to adversaries because it is a single and independent node that can be attacked easily. For the problem of availability Anderson et al. [4] suggested an approach in which CAs are assigned to the nodes repeatedly. While it seems that this technique may solve the availability problem (because the network will be working well as long as there is just one node in the MANET), this approach may become unstable when the nodes are trying to find each other in the network. One solution is to utilize a Distributed Certificate Authority (DCA). In section 2 the concept of DCAs in MANETs is introduced. In section 3 Threshold Cryptography is described and section 4 compares and categorizes different types of DCAs. Finally, section 5 proposes the best DCA system for MANETs.

## 2. Distributed Certificate Authority

A Distributed Certificate Authority (DCA) system is defined in a situation that the private key of the CAs are distributed among the network's nodes. All of the nodes in the MANET will have the public key of the CAs for verifying the signatures signed by CAs (i.e. shareholders that have to participate in issuing and verifying signatures [5]). A threshold is defined for the maximum number of shareholders in the procedure. Table 1 compares the difference between a DCA and the more traditional Centralized Certificate Authority (CCA). The table shows that adopting a distributed mode will affect the level of security, availability and reliability.

**Table 1.** Comparison between CCA and DCA

|                         | CCA  | DCA  |
|-------------------------|------|------|
| Availability            | LOW  | HIGH |
| Security                | HIGH | LOW  |
| Performance             | HIGH | LOW  |
| Scalability             | HIGH | LOW  |
| User Mobility           | HIGH | ---  |
| DCA Mobility            | LOW  | HIGH |
| Validity of Certificate | HIGH | LOW  |

There are many DCAs that have been designed for MANETs and these can be categorized into two groups: Partially Distributed Certificate Authority (PDCA) and Fully Distributed Certificate Authority (FDCA).

In FDCA all the nodes are shareholders and can generate certificates. FDCA has the potential to be attacked and broken, since one intruder who can attack one or more nodes can enter the network. As Dhillon et al. [6] proposed this problem can be overcome by providing a powerful Intrusion Detection System (IDS) to identify the compromised nodes. Moreover, a limited life time can be defined for the certificates so that after expiration the certificates cannot be used. It is necessary to decide between security and performance to choose a suitable expiration time. If large expiration times for certificates are chosen, security will weaken however if the expiration times are frequently renewed, it will cause a large amount of data transfer in the network that will result in over-heading.

In an FDCA all the nodes in the network share the secret, while in PDCA, some specific nodes are generating certificates and a node can combine a small number of these shares to produce a valid certificate. It is necessary to have a powerful server in this approach that is responsible for selecting the nodes for secret sharing. Both approaches suffer from disadvantages with a key one being availability. It is difficult to ensure that all the nodes assigned for sharing the secret are available at a given time. There are also problems of performance and ensuring the suitability of nodes which are affected by many factors such as the scale of the network, level of security and also the architecture of the network.

**Table 2.** Comparison between PDCA & FDCA

|                 | PDCA             | FDCA      |
|-----------------|------------------|-----------|
| Security        | Higher than FDCA | LOW       |
| Availability    | Lower than FDCA  | HIGH      |
| Scalability     | HIGH             | LOW       |
| Mobility Support| LOW              | HIGH      |
| Network Size    | Large            | Small     |
| IDS Monitoring  | Not required     | Required  |
| Secret Updates  | Multicast        | Broadcast |

## 3. Secret Sharing

In a Distributed Certificate Authority, actions like generating a digital signature and certificates are distributed through the limited number of nodes so that the selected nodes are participating in doing these works. In Threshold Cryptography (TC) a (k, n) threshold is used that allows a CA's certificate to be divided into n splits, where n is the total number of subgroup members and k<n shareholders can find the certificate by combining their own shared key that k-1 or fewer shareholders cannot do so [7]. In this approach, if the attacker could find the shared secret of few shareholders but less than k, they cannot recover the certificate. However, this approach will not work well if the attacker has found more than k. For this reason the secret sharing has to be updated by providing new set of shares periodically [8].

## 4. Secret Share Updating

If attackers could find k shareholder and compromise them in a certain time, they could enter the whole network. So rather than shared secret splitting algorithms that are supposed to be secure, it seems that there is a need to update the shareholders in certain time intervals. In this case it is not necessary to change the secret key itself. The attackers must compromise k shareholders between the time intervals. Specifying update intervals has an effect on the security and performance of the network. Thus, if shorter times are chosen, it may result in an over-head on the network and also choosing a long time interval may reduce the security [9].

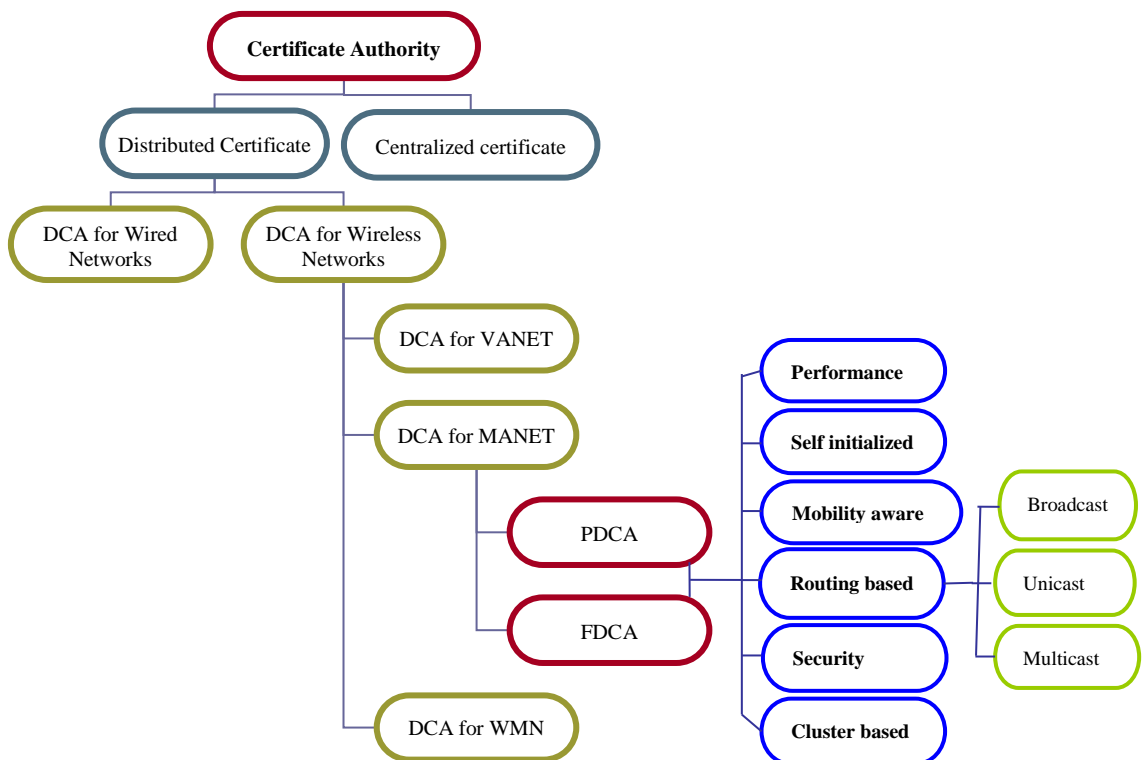

**Figure 1**. Distributed Certificate Authority Hierarchy.

## 5. Distributed Certificate Authority Categories

In Figure 1, different types of DCAs are categorized into 6. Because ad hoc networks are not very flexible in scalability and performance, Clustering can help to overcome these limitations. It can reduce the file storage in the nodes which are going to have the certificates of other nodes in clusters, rather than the whole network. Also, it can reduce the overall load of the network by dividing nodes into groups and provide more efficient certificate management.

**A.   Clustering based DCA schemes**

Chaddoud et al. [5] suggested a cluster based DCA in which certificates are distributed through Cluster Heads (CH) which are shareholders in the network. This ensures that none of the single cluster heads know the DCA and when a new cluster head joins the network, it has to be signed with the shared private key. In this case the new node sends its request to get the DCA's share. Any cluster head which receives the request signs the key and provide a share for the new node. As soon as the node receives its shared key, it can get the full certificate on demand.

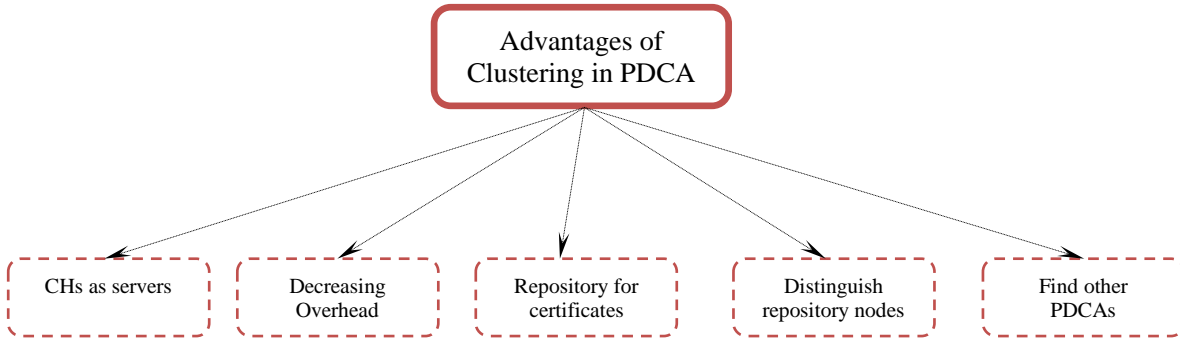

**Figure 2**. Advantages of Clustering in PDCA.

There is another cluster-based DCA which is proposed by Rao et al. [10] that categorizes the network's nodes into 3 groups: Repository nodes; Client nodes; and, server nodes. In this scheme whole nodes are organized into clusters. From each cluster, some nodes are selected to be repository nodes and after that the server nodes are selected through repository nodes. When a new node enters the network, it has to contact a Registration Authority section. The Registration Authority section sends a request to the server nodes. The server nodes sign the certificate and resend it to the Registration Authority section and at the end the signed certificate will be given to the new node by the Registration Authority section.

The problem with this approach is that the registration authority section was part of another wired network that was working dependently. However, the mobility of the nodes has been considered in this scheme so that they could handle any topology changes in the MANET [11].

The third scheme was proposed by Elhdhili et al. [12] in which the DCA is based on clustering, is totally distributed and uses a (k, n) threshold, RSA signed certificate to the group of cluster heads. In this approach there are 3 types of nodes which are: Administrator; Cluster Head; and, Cluster members.

Another approach which is proposed by Lee et al. [13], is a partially distributed certificate authority that supports node mobility and movements. This approach is more scalable as it allows the nodes to authenticate each other. In this model creating certificates also takes less time despite the increased packet sizes. It is considered that the certificate creation procedure is not affected by the number of nodes that exist in the network, as the certificates are produced by the existing members and are not interrupted by nodes joining the cluster.

**Table 3.** Properties of Cluster-based DCA

| Node Types | CA storage | Security | Authentication |
| --- | --- | --- | --- |
| Cluster & CH | - | CH share Updates | - |
| Client, Repository, Server | Repository Nodes | Revocation by CRL | Registration Authority |
| Administrator, CH, regular | Administrator Nodes | Secure inside cluster | Nodes Participation |

**B.   Routing-based DCA**

The easiest way to send certificate messages through the whole of the network is by using broadcasting. However, this way of sending messages causes an increased over-head and reduces the MANET's performance, so the recommended solution is to use Unicast protocols for DCA approaches. Unicast DCA protocols can be classified into three groups: reactive; proactive; and, hybrid protocols.

In one of the routing based DCAs proposed by Xia et al. [14] identity-based FDCA is used in a MANET which is more adequate for MANETs because it can reduce the overhead in the network. This model is based on proactive routing, however Sen et al. [15] suggested the Mobile Certificate Authority (MOCA) protocol which was more reliable and efficient than the original MOCA presented by Rao et al. [10].

**Table 4.** Properties of Routing based DCA

| Routing protocol | Security | Optimization |
|---|---|---|
| Proactive Routing | Utilize route cache | Use Unicast |
| Reactive Routing | - | Change routing packets |

C. Self-Initializing Protocol

One of the main issues in MANETs is the initialization and startup procedures. A self-initialized system is needed to begin its security related works on system boot time and to provide certificate authorities by using Self-Initializing Protocols (SIP). Ge et al. [16], proposed a self-initialized DCA which is more scalable, cost efficient and more secure. In this method all the factors and parameters required for DCA such as total members and threshold values will be defined.

Kang et al. [17] suggested another scheme of Self-Initialized DCA (SDCA) in which the nodes that can distribute partial keys, are authenticated by a system authority section.

D.   Mobility Supported Schemes

As it is critical to have enough nodes for creating certificates, the mobility and existence of the nodes will have an effect on a DCAs operations. Some methods have been proposed to be aware of the mobile nodes as described in the following sections.

Pereira et al. [18] suggested a new mobility aware scheme in which a DCA system can adapt itself with the members' existence and also provides availability and reliability for the DCA system. Joshi et al. [19] proposed a scheme that produces more shares for each node. As a result, certificates can be recreated with a smaller number of existing nodes.

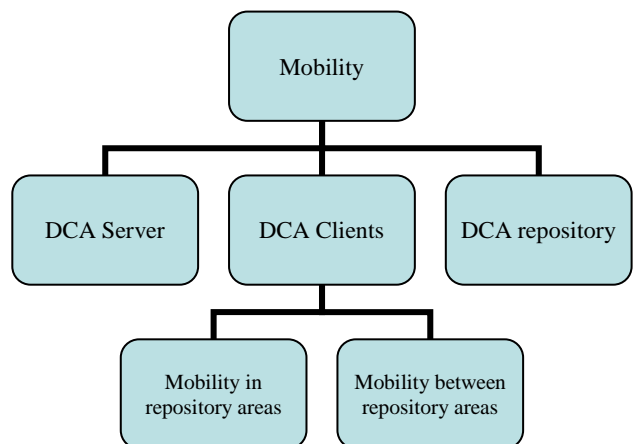

**Figure 3**. Mobility Schemes Hierarchy.

E.   Security-aware Schemes

From a security point of view, there are some proposed DCA systems which are secure against various type of attacks in MANETs. Zhou et al. [20] proposed a method that is based on multiple key cryptography DCA system. Also, Rajaram et al. [21], Zeb et al. [22], Dhabi et al. [23] and Chaudhry et al. [24] suggested a powerful scheme that supports certificate updates for preventing various attacks. Figure 4 present techniques that can help to make DCA system more secure.

**6. Revised DCA System**

After studying the properties of an efficient MANET Certificate Authority there are some important factors to define in order to establish a powerful, secure and efficient DCA system for MANETs. Chaddoud et al. [5] have suggested a scheme for DCA system. The following sections review these factors and add important issues which need to be considered in designing such systems.

- Availability

All the nodes in the network which are shareholders need to be available in the MANET. A good and efficient MANET should have solutions for the mobility and availability of the nodes and ensures that there are always enough shareholders to produce certificates.

- Reliability

MANETs by their nature are unreliable because of node mobility and their wireless communications.

- Security

One of the most important security issues in a MANET is avoiding a node acting as a single point of failure. For this purpose, secret sharing algorithms are used as well as certificate updating.

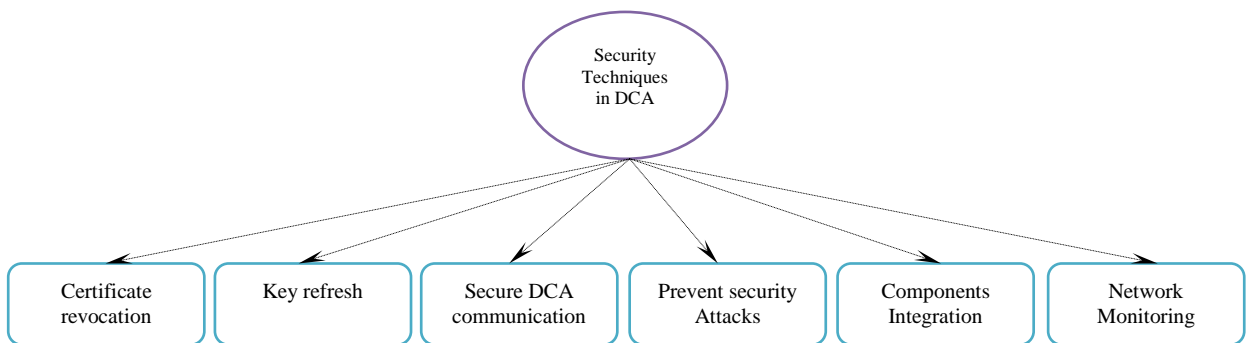

**Figure 4**. Security Techniques in DCA.

- Efficiency

Some of the MANETs' limitations are bandwidth, scalability and data communications through wireless media. These factors need to be carefully studied to design a powerful DCA system.

- Fault tolerant

A well-designed DCA system needs to ensure that all the components in the MANETs are continuing their own correct predefined tasks. There is also a need to have some monitoring control systems to detect any fault through the entire network.

- Node Mobility

There are some kinds of mobility in ad hoc networks that should be covered by a DCA system. One of these is a clients' mobility inside their own clusters and even from one cluster to another one. Another mobility is the repository nodes' mobility in the network or between two or more networks.

- Self-initialization

This part can be studied from two points. The first one is designing an automatic system which can support all DCA tasks and another one is to have a self-initialization system which is going to make the DCA work well on network start up.

- Coordination with network and integration

A DCA system which is implemented on top of an ad hoc network needs to be adapted well with all the protocols used in wireless networking and especially in ad hoc networks.

- Scalability

The reliability and security of the networks will be reduced with the MANETs expansion. But there are some solutions that can be used to scale a DCA systems in MANETs without facing any problem or at least with fewer limitations.

- Independence

MANETs, like any other topologies, need to be independent from other infrastructure like wired networks because any reliance in distributed topologies like ad hoc networks will cause some problems.

•   Storage efficiency

Public key infrastructure needs a large amount of space for encryption and decryption, so it is very important to choose a suitable data structures to prevent any other problems caused by shortage of space.

## 7. Conclusion

With security playing a vital role in MANETs, there are many proposed methods to prevent attacks. One of the main security issues in MANETs is the Certificate Authority service. By using PKI it is possible to design a secure and efficient ad hoc network which is at least as secure as wired networks. In this paper different distributed certificate authorities were proposed to adapt PKI components for use in wireless networks. This classification enables a better understanding of the concepts and the identification of solutions for unsupported areas or less concentrated issues.